\documentclass[aps,prl,preprint,showpacs]{revtex4}
\usepackage{bm}
\usepackage{graphicx}
\begin{document}
\title{Breakdown of fluctuation-dissipation relations in granular gases}
\author{J. Javier Brey}
\email{brey@us.es}
\author{M.I. Garc\'{\i}a de Soria, and P. Maynar}
\affiliation{F\'{\i}sica Te\'{o}rica, Universidad de Sevilla, Apdo.\
de Correos 1065, E-41080 Sevilla, Spain}

\date{today}

\begin{abstract}
A numerical molecular dynamics experiment measuring the two-time
correlation function of the transversal velocity field in the
homogeneous cooling state of a granular gas is reported. By
measuring the decay rate  and the amplitude of the correlations, the
accuracy of the Landau-Langevin equation of fluctuating
hydrodynamics is checked. The results indicate that although a
Langevin approach can be valid, the fluctuation-dissipation relation
must be modified, since the viscosity parameter appearing in it
differs from the usual hydrodynamic shear viscosity .
\end{abstract}

\pacs{45.70.-n,45.70.Mg,51.10.+y,05.20.Dd}

\maketitle

The hydrodynamic description of ordinary one-component fluids is
made by means of the local number density, $n({\bm r},t)$, the flow
velocity, ${\bm u}({\bm r},t)$, and the temperature, $T({\bm r},t)$
\cite{LyL59}. These fields are determined from some closed
macroscopic equations, e.g. the Navier-Stokes equations, with the
appropriate initial and boundary conditions. On the other hand, the
most detailed microscopic description of the evolution of the fluid,
in terms of the properties of the particles forming it, is given by
non-equilibrium statistical mechanics. In this latter formulation,
the hydrodynamic fields appear as the ensemble average of some
associated microscopic fields, being also possible to describe their
stochastic properties, at least formally. There is an intermediate
level of description, often referred to as the mesoscopic
description, in which macroscopically fluctuating hydrodynamic
fields are considered.

The first formulation of fluctuating hydrodynamics for equilibrium
states was proposed by Landau and Lifshitz \cite{LyL59}. Their
equations describe the fluctuations of the hydrodynamic fields about
their equilibrium values, and have the form of linear Langevin
equations. The noise terms are taken as white and Gaussian with the
second moments determined by the Navier-Stokes transport
coefficients of the fluid. This establishes a relationship between
the fluctuations of the fields at equilibrium and the dissipation by
transport in the fluid, and so are known as fluctuation-dissipation
(F-D) relations of the second kind \cite{KTyN85}. Since the
formulation of the theory, the question of whether the same idea can
be used to describe fluctuations in systems that are far away from
equilibrium has attracted the attention of scientists. For molecular
fluids, it seems well established that the F-D relations apply under
the same conditions as the usual Navier-Stokes hydrodynamic
equations \cite{OyS06}.

Granular fluids are always out of equilibrium due to the kinetic
energy dissipation in collisions. Evidence has been accumulated
during the last years that they provide a unique proving ground for
non-equilibrium statistical mechanics ideas and techniques.
Nevertheless, the analogy between granular and ordinary fluids must
not be pushed too far. Granular materials have many peculiar
properties arising from the macroscopic size of the particles and
the lack of kinetic energy conservation before and after a collision
\cite{JNyB96}. In particular, the validity of the F-D relations can
not be assumed {\em a priori}.

Here a definite test of the validity of the F-D relations for
granular gases is carried out under the most favorable and ideal
conditions: the simplest model, state, and hydrodynamic field. If
the F-D fails in this case, it will also certainly fail for more
``realistic'' conditions. More precisely, the behavior of the
transversal component of the flow field in the so-called homogeneous
cooling state (HCS) of a granular gas is the topic addressed in this
paper. Conceptually, this problem is similar to the experimental
situation investigated in ref. \cite{AMBLyN03}, although the
granular state considered is different. To start with, the
Landau-Langevin formalism for the transversal velocity field in an
equilibrium ordinary fluid will be briefly recalled.

Consider the local fluctuations of the velocity flow $\delta {\bm u}
({\bm r},t)$ around the equilibrium value ${\bm u}_{e}={\bm 0}$, and
introduce its Fourier representation $\delta {\bm u}_{\bm q}(t)$.
This vector can be decomposed into the component parallel to ${\bm
q}$ and another component perpendicular to it: $\delta {\bm u}_{\bm
q} (t)= \delta u_{{\bm q}, \parallel} \widehat {\bm q} + \delta {\bm
u}_{{\bm q}, \perp} (t)$, with $\widehat{\bm q} \equiv {\bm q}/q$
and $\delta {\bm u}_{{\bm q}, \perp} \cdot \widehat{\bm q}=0$. The
vector component $\delta {\bm u}_{{\bm q}, \perp} (t)$ is the
vorticity or transversal flow field and obeys the Langevin equation
\cite{LyL59}
\begin{equation}
\label{1} \left[ \partial_{t}+ \left( m n_{e} \right)^{-1} \eta_{e}
q^{2} \right] \delta {\bm u}_{{\bm q}, \perp} (t) = {\bm f}_{{\bm
q},\perp}(t),
\end{equation}
where $m$ is the mass of the fluid particles, $\eta_{e}$ the
Navier-Stokes shear viscosity of the fluid, and ${\bm f}_{{\bm q},
\perp} (t)$ a randomly fluctuating force that is a Gaussian white
noise of zero average. It verifies the fluctuation-dissipation
relation
\begin{equation}
\label{2} \langle {\bm f}_{{\bm q}, \perp} (t) {\bm f}_{{\bm
q}^{\prime}, \perp} (t^{\prime}) \rangle_{e} =
\frac{2T_{e}V}{m^{2}n_{e}^{2}}\, \delta (t-t^{\prime}) \delta_{{\bm
q},-{\bm q}^{\prime}} \overline{\eta}_{e} q^{2} {\sf I}.
\end{equation}
In the above expressions, angular brackets denote average over the
noise realizations, the index {\em e} is used to indicate quantities
evaluated at equilibrium, $V$ is the volume of the system, ${\sf I}$
is the unit tensor in the subspace perpendicular to ${\bm q}$, and
$\overline{\eta}_{e}$ a coefficient that according with the Landau
theory is the same as the viscosity, $\overline{\eta}_{e}
=\eta_{e}$. This is a main physical assertion of the theory.

There is no equilibrium state for granular fluids. Instead, the
simplest state for a freely evolving granular fluid is the HCS,
mentioned above. At a macroscopic level, it is characterized by a
uniform density $n_{H}$, a vanishing flow velocity, and a
monotonically decreasing granular temperature, $T_{H}(t)$, that
obeys the Haff law \cite{Ha83},
\begin{equation}
\label{3}
\partial_{t} T_{H}(t) = - \zeta_{H}(T_{H}) T_{H}(t),
\end{equation}
$\zeta_{H}(T_{H})$ being the cooling rate accounting for the energy
dissipation in collisions. The question addressed here is whether
Eqs. (\ref{1}) and (\ref{2}) are also valid to describe fluctuations
of the vorticity in the HCS of a granular fluid as, in fact, has
been assumed sometimes in the literature \cite{vNEByO97}. Of course,
if the answer is negative, there is little hope that those equations
hold for stronger non-equilibrium states, such as vibrofluidized
granular media.

The simplest and most used model for granular fluids is an ensemble
of $N$ smooth  inelastic hard spheres ($d=3$) or disks ($d=2$) of
diameter $\sigma$ \cite{Go03}. Inelasticity is described by means of
a constant coefficient of normal restitution $\alpha$, defined in
the interval $0 < \alpha \leq 1$. For this model, expressions for
the transport coefficients have been obtained by using the inelastic
Boltzmann-Enskog equation and the Chapman-Enskog procedure. In
particular it has been found that \cite{BDKyS98,GyD99}
\begin{equation}
\label{4} \eta_{H}=m n_{H} \ell_{0} v_{H}(t) \eta^{*}(\alpha,n_{H}),
\end{equation}
\begin{equation}
\label{5} \zeta_{H}(T_{H})= v_{H}(t) \ell_{0}^{-1}
\zeta^{*}(\alpha,n_{H}),
\end{equation}
where $\ell_{0} \equiv (n_{H} \sigma^{d-1})^{-1}$ is proportional to
the mean free path of the particles, $v_{H}(t) \equiv \left[ 2
T_{H}(t) /m \right]^{1/2}$ is a thermal velocity for the HCS, and
$\eta^{*}$ and $\zeta^{*}$ are dimensionless functions of the
density and the coefficient of restitution. When Eqs. (\ref{1}) and
(\ref{2}) are applied to the HCS, the problem arises that the
viscosity and the second moment of the noise term are time-dependent
through the temperature $T_{H}(t)$. It is then convenient to
introduce dimensionless length and time scales by ${\bm \ell} \equiv
{\bm r} /\ell_{0}$ and $ds \equiv v_{H}(t) dt /\ell_{0}$,
respectively. Moreover, a dimensionless velocity field is defined by
${\bm \omega} ({\bm l}, s) \equiv \delta {\bm u}({\bm r},t)
/v_{H}(t)$. Then, Eqs. (\ref{1}) and (\ref{2}) become
\begin{equation}
\label{6} \left( \partial_{s} - \zeta^{*}/2 +\eta^{*} k^{2} \right)
{\bm \omega}_{{\bm k}, \perp} (s) = {\bm \xi}_{{\bm k}, \perp} (s),
\end{equation}
\begin{equation}
\label{7} \langle {\bm \xi}_{{\bm k},\perp} (s) {\bm \xi}_{{\bm
k}^{\prime}, \perp}(s^{\prime}) \rangle_{H} = V^{* 2} N^{-1} \delta
(s-s^{\prime}) \delta_{{\bm k}, -{\bm k}^{\prime}}
\overline{\eta}^{*} k^{2} {\sf I},
\end{equation}
with ${\bm k} = \ell_{0} {\bm q}$ and $V^{*}= V/\ell_{o}^{d}$.
Moreover, in Eq. (\ref{7}) it is $\overline{\eta}^{*} = \eta^{*}$.
For wavevectors ${\bm k}$ such that $\lambda_{\perp} (k) \equiv
\zeta^{*}/2- \eta^{*} k^{2} <0$, the long time solution of Eq.
(\ref{6}) reads
\begin{equation}
\label{8} {\bm \omega}_{{\bm k}, \perp} (s) = \int_{- \infty}^{s}
ds_{1}\, e^{\lambda_{\perp}(k) (s-s_{1})} {\bm \xi}_{{\bm k}, \perp}
(s_{1}).
\end{equation}
Using the above expression and Eq.\ (\ref{7}), the two-time
correlation function of the transverse velocity field is easily
obtained,
\begin{equation}
\label{9} {\sf C}_{{\bm k},{\bm k}^{\prime}}(s,s^{\prime}) \equiv
\langle {\bm \omega}_{{\bm k},\perp}(s) {\bm \omega}_{{\bm
k}^{\prime},\perp}(s^{\prime}) \rangle_{H} = - \frac{V^{*2}
\overline{\eta}^{*}k^{2}}{2N \lambda_{\perp}(k)}\, \delta_{{\bm
k},-{\bm k}^{\prime}} e^{\lambda_{\perp}(k)(s-s^{\prime})} {\sf
I},
\end{equation}
valid for $s \geq s^{\prime} \gg 1$.

\begin{figure}
\includegraphics[scale=0.5,angle=0]{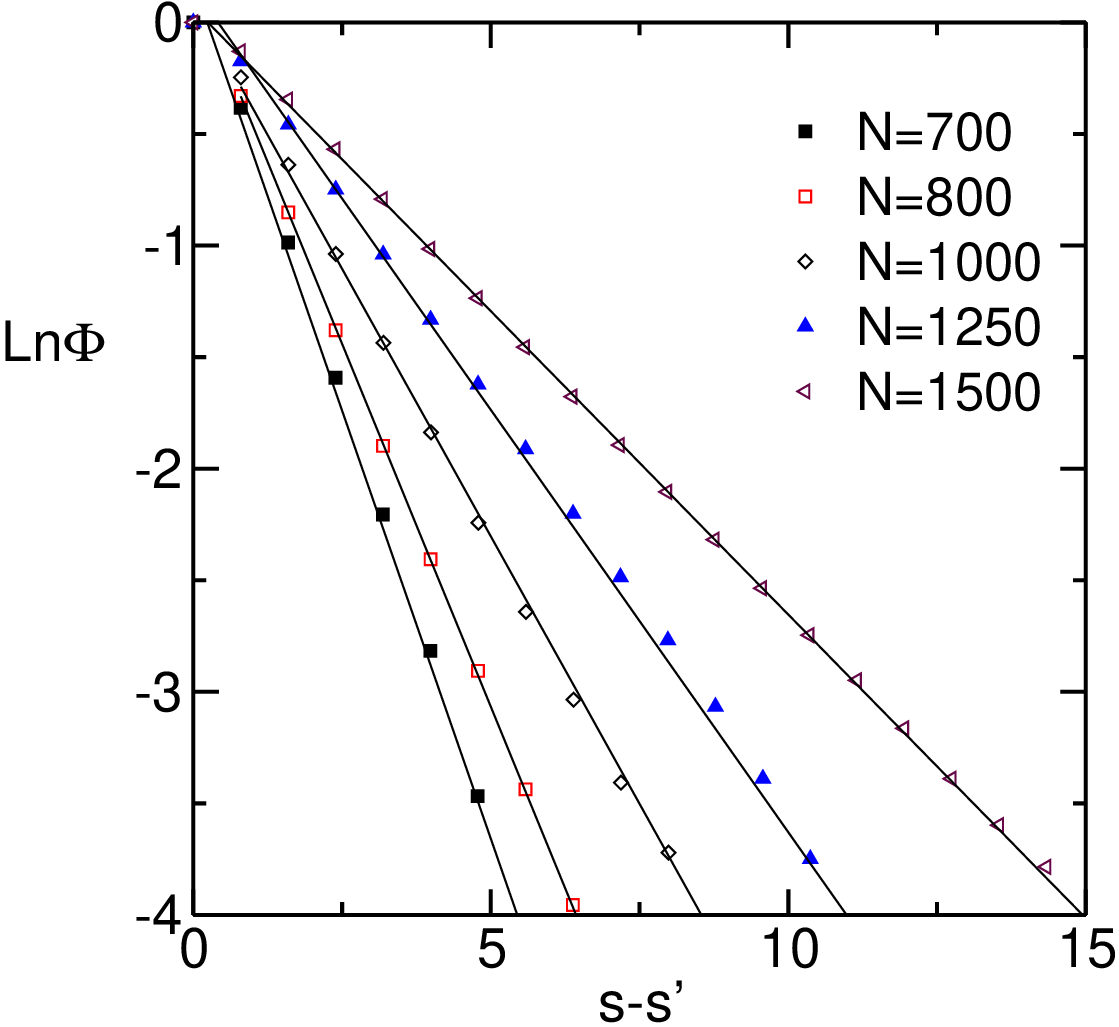}
\caption{MD simulation results showing the exponential decay of the
two-time correlation function of the transversal velocity for
$\alpha=0.8$ and $n_{H}\sigma^{2}=0.01$. \label{fig1}}
\end{figure}

\begin{figure}
\includegraphics[scale=0.5,angle=0]{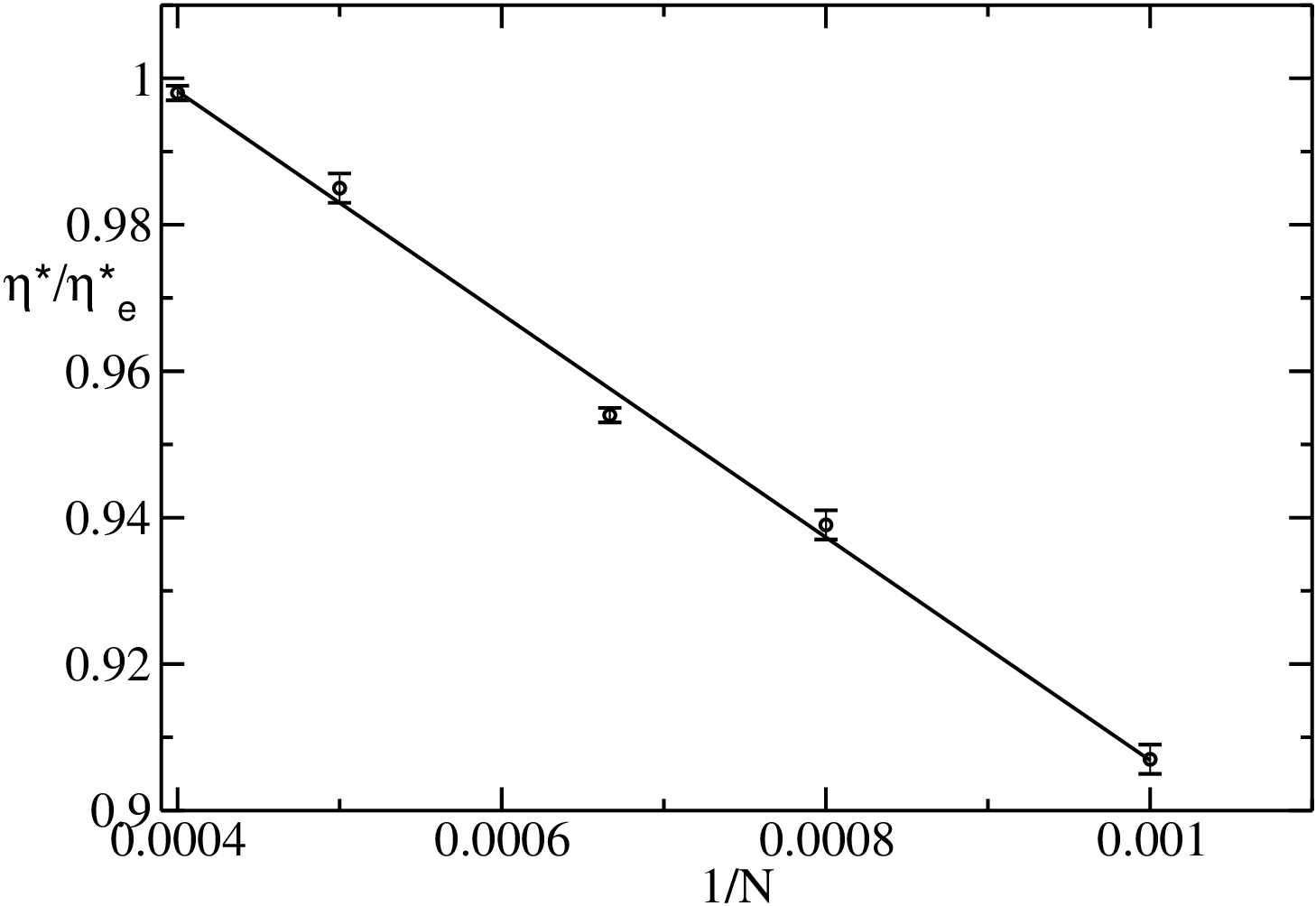}
\caption{Values of the parameter $\eta^{*}/\eta^{*}_{e} = 2
\sqrt{2 \pi} \eta^{*}$, obtained from the exponential decay of the
correlation function of the transversal velocity field, as a
function of the inverse of the number of particles $N$. The
density and coefficient of restitution are the same as in Fig.
\ref{fig1}. Symbols are from MD simulations and the solid line is
the best linear fit. \label{fig2}}
\end{figure}

In the following, MD simulation results testing Eq. (\ref{9}) will
be reported. Consider a square cell ($d=2$) of side $L$ with
periodic boundary conditions, so the minimum value of $k$ allowed is
$k_{min}= 2 \pi \ell_{0}/L$. The value of $L$ has been always chosen
such that $k_{min} > k_{\perp} \equiv \left( \zeta^{*}/2
\eta^{*}\right)^{1/2}$, to guarantee that the system be always in
the parameter region in which it is expected to be stable with
respect to vorticity fluctuations, i.e. $\lambda_{\perp} (k_{min})
<0$ and Eq.\ (\ref{9}) holds for all the allowed values of $k$. In
the simulations, fluctuations corresponding to ${\bm k} = k_{min}
\widehat{\bm e}_{x}$, where $\widehat{\bm e}_{x}$ is the unit vector
in the direction of one of the sides of the cell, have been
measured. Therefore, ${\bm \omega}_{{\bm k},\perp} = \omega_{{\bm
k},\perp} \widehat{\bm e}_{y}$, with $\widehat{\bm e}_{y}$ being an
unit vector perpendicular to $\widehat{\bm e}_{x}$. The particle
representation of the transversal velocity field considered here is
\begin{equation}
\label{10} \delta u_{q_{min}\widehat{\bm e}_{x},\perp}(t) =
\sum_{i=1}^{N} e^{-2 \pi i X_{i}(t)/L} V_{y,i}(t),
\end{equation}
where the components of the position and velocity of particle $i$ at
time $t$ have been denoted by $X_{i}(t),Y_{i}(t)$ and
$V_{x,i}(t),V_{y,i}(t)$, respectively. In the MD simulations an
event driven algorithm and the steady state representation of the
HCS \cite{Lu01,BRyM04} have been employed. The fluctuations of the
transverse velocity turned out to be always Gaussian within the
statistical errors. Moreover, the imaginary part of the two-time
correlation function was found to vanish, as expected by symmetry
considerations and in agreement with Eq.\ ({\ref9}). The restriction
to the parameter region in which the HCS is stable implies that the
size of the system and, therefore, the number of particles at a
given density, is bounded from above. Consequently, relevant size
effects show up at low densities. In Fig. \ref{fig1},
\begin{equation}
\label{11} \Phi(s,s^{\prime}) \equiv \frac{C_{k_{min}\widehat{\bm
e}_{x},-k_{min}\widehat{\bm e}_{x}}(s,s^{\prime})}
{C_{k_{min}\widehat{\bm e}_{x},-k_{min}\widehat{\bm
e}_{x}}(s^{\prime},s^{\prime})}
\end{equation}
is plotted as a function of $s-s^{\prime}$ for $\alpha=0.8$, $n_{H}
\sigma^{2}=0.01$, and several values of $N$, as indicated. In all
cases, an exponential decay is observed, in agreement with Eq.
(\ref{9}), that predicts $\Phi(s,s^{\prime})= \exp \left[
\lambda_{\perp}(k_{\min}) (s-s^{\prime}) \right]$. From the data,
values for $\lambda_{\perp}(k_{min})$ are obtained. Of course,
similar results are found when the labels of the two axis are
interchanged. The simulations also provide the value of the cooling
rate, that is independent of $N$ and in good agreement with the
theoretical prediction given in \cite{BDKyS98,GyD99}. Then, by
subtracting $\zeta^{*}/2$ from the obtained values of
$\lambda_{\perp}(k_{min})$,  simulation results for the parameter
$\eta^{*}$ follow. From the analysis of the data, it follows that
the value obtained depends on the number of particles $N$ (size $L$)
employed in the simulation. The theoretical prediction corresponds
to the formal limit $N \rightarrow \infty$, $L \rightarrow \infty$,
with  $n_{H}$ constant. In Fig. \ref{fig2}, the values of $\eta^{*}$
obtained from the data in Fig. \ref{fig1} are plotted as a function
of $N^{-1}$. A linear fit leads to the result $\lim_{N \rightarrow
\infty} \eta^{*} = (1.059 \pm 0.004) \eta^{*}_{e}$, where
$\eta^{*}_{e}=  (2 \sqrt{2 \pi})^{-1}$ is the low density
(Boltzmann) elastic limit. The theoretical prediction for the
Navier-Stokes shear viscosity obtained from the Boltzmann equation
in the first Sonine approximation is $\eta^{*} = 1.057
\eta^{*}_{e}$. Moreover, if a ``size-dependent viscosity'' is
measured by considering the decay of a small macroscopic sine
perturbation of the velocity field \cite{BRyC99} in systems of
different sizes, the values are in good agreement with those
reported in Fig. \ref{fig2}. Therefore, it can be concluded that,
for the system being investigated, Eq. (\ref{9}) predicts accurately
the decay of the correlation function, with $\eta^{*}$ being the
(inelastic) Navier-Stokes shear viscosity.

\begin{figure}
\includegraphics[scale=0.5,angle=0]{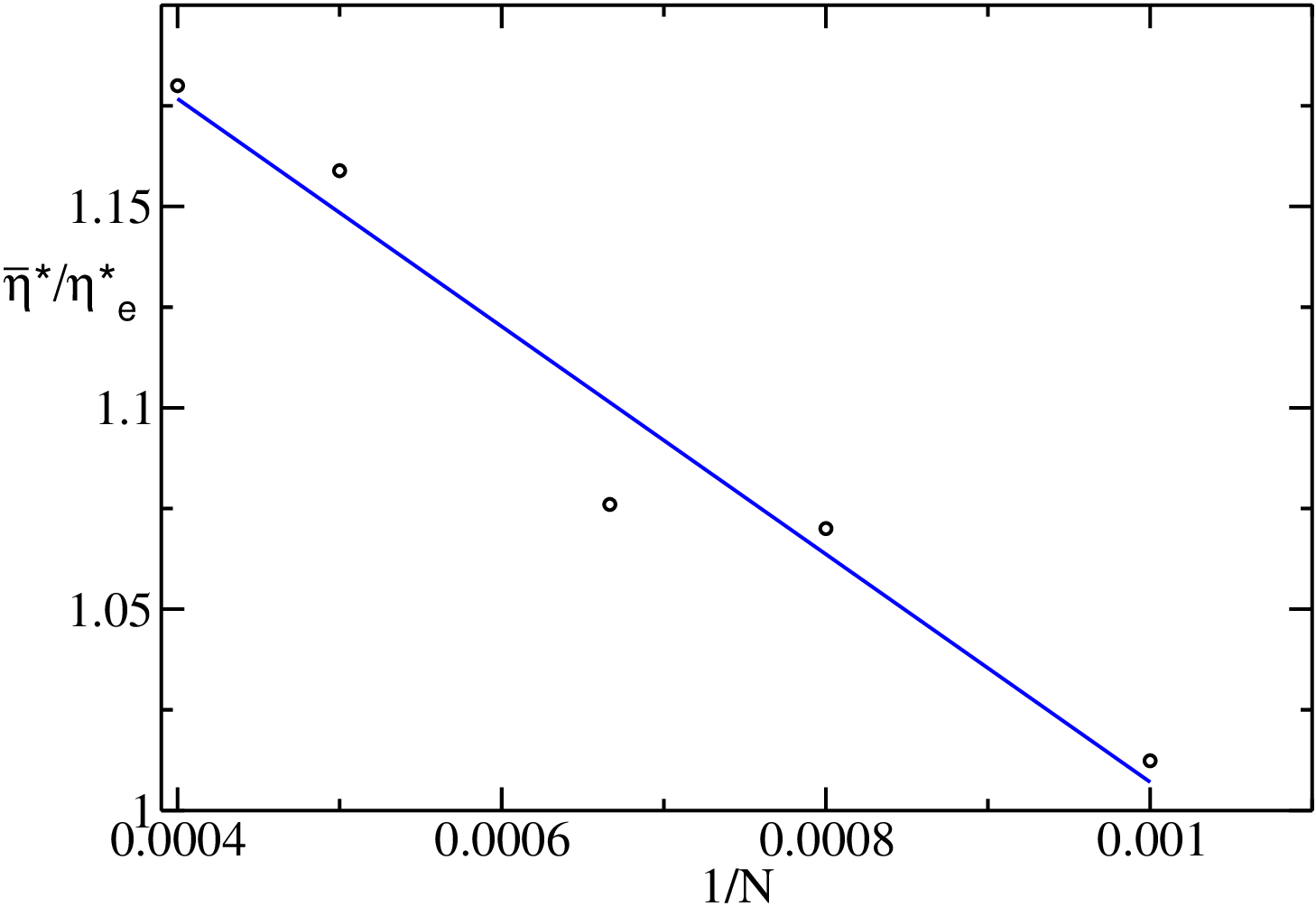}
\caption{Values of the parameter $\overline{\eta}^{*}/\eta^{*}_{e} =
2 \sqrt{2 \pi} \overline{\eta}^{*}$, obtained from the amplitude of
the fluctuations of the transversal velocity field, as a function of
the inverse of the number of particles. The density and coefficient
of restitution are the same as in Fig. \ref{fig1}. Symbols are from
MD simulations and the solid line is the best linear fit.
\label{fig3}}
\end{figure}

Focus next on the amplitude of the fluctuations. From Eq.
(\ref{9}), it follows that
\begin{equation}
\label{12} C_{k_{min}\widehat{\bm e}_{x},- k_{min}\widehat{\bm
e}_{x}} (s,s) = - \frac{V^{*2} \overline{\eta}^{*} k_{min}^{2}}{2 N
\lambda_{\perp}(k_{min})}\, .
\end{equation}
Since $\lambda_{\perp}(k_{min})$ was determined from the time decay
of the correlation function, the coefficient $\overline{\eta}^{*}$
can be computed from the measurements of the above quantity. Again,
a clear dependence on the number of particles $N$ is observed in the
results reported in Fig.\ \ref{fig3}, that correspond to the same
system as in Figs. \ref{fig1} and \ref{fig2}. By means of a linear
fit, the asymptotic $N \rightarrow \infty$ value
$\overline{\eta}^{*} = 1.29 \pm 0.02$ is found. It differs from both
the shear viscosity of the system measured by the procedures
described above and the theoretical prediction in more than $20 \%$.

\begin{figure}
\includegraphics[scale=0.5,angle=0]{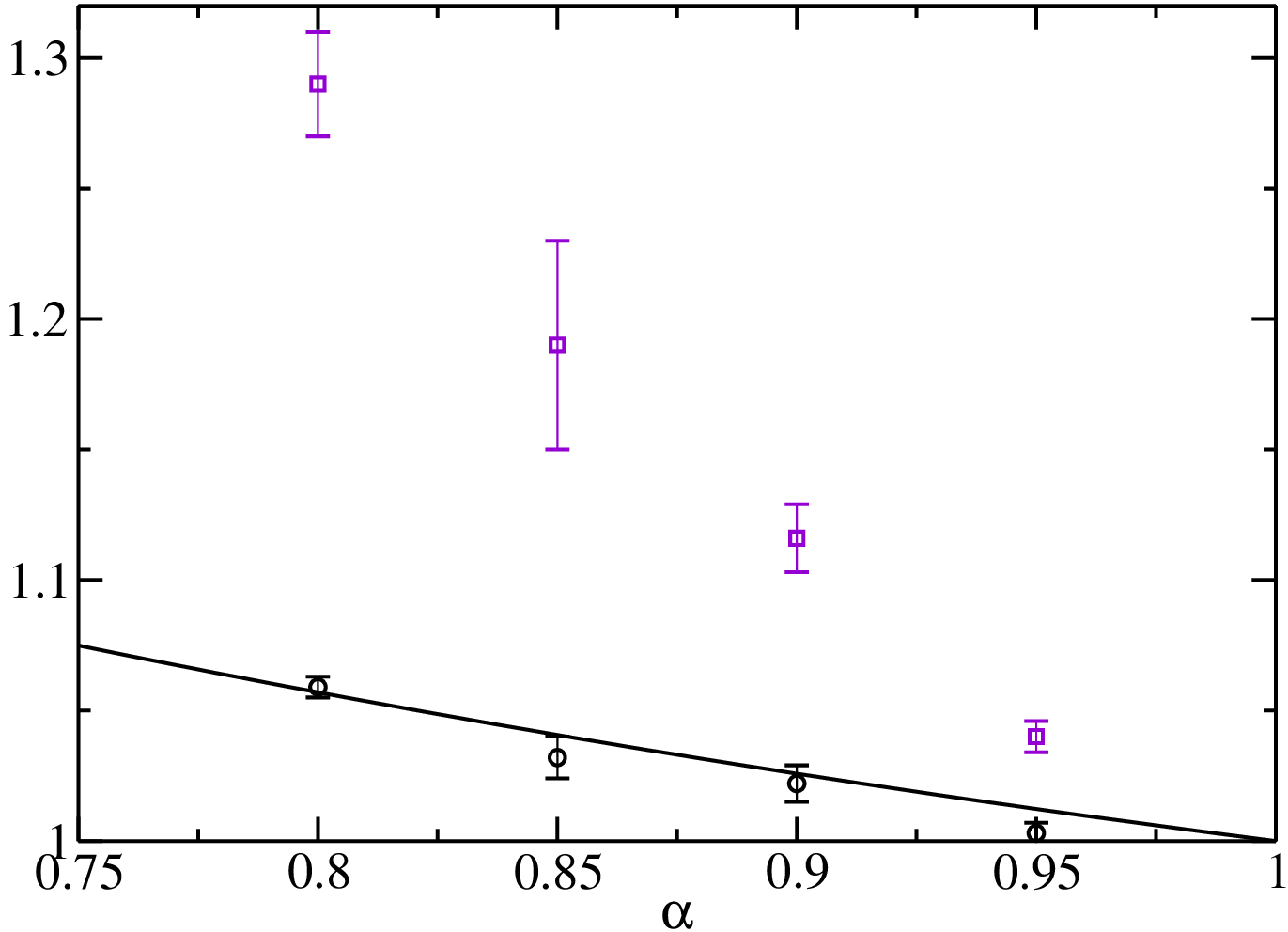}
\caption{Coefficients $\eta^{*}/\eta^{*}_{e}$ (circles) and
$\overline\eta^{*}/\eta^{*}_{e}$ (squares) measured from the MD
simulations and extrapolated to $N \rightarrow \infty$, as functions
of the coefficient of restitution $\alpha$. The density is always
very small. The solid line is the Chapman-Enskog prediction in the
first Sonine approximation. \label{fig4}}
\end{figure}

The above analysis has been repeated for different values of the
coefficient of restitution, keeping the density small to allow
comparison with the kinetic theory predictions. In particular, the
results plotted in Fig. \ref{fig4} correspond to the case previously
discussed and to $\alpha=0.85$, $0.90$, and $0.95$, with $n_{H}
\sigma^{2}=6 \times 10^{-3}$, $0.018$, and $0.02$, respectively.  It
is observed that the discrepancy between the shear viscosity $\eta$
and the coefficient $\overline{\eta}$ characterizing the second
moment of the fluctuations, increases as the inelasticity increases,
vanishing in the elastic limit, as required. In the simulations,
attention has been paid to assure that $k_{min}$ is neither too
large nor too small. In the former case, the Navier-Stokes
approximation is expected to fail, while in the latter instability
effects show up.

Two main physical conclusions follow from the study presented here.
The Landau-Langevin  fluctuating hydrodynamic equations cannot be
directly extended to granular fluids. But, on the other hand, it
seems possible to separate slow and fast degrees of freedom and to
model granular systems in the spirit of the Langevin approach. A
similar conclusion was reached from the study of the fluctuations of
the total energy of the system \cite{BGMyR04}. The open and relevant
question is how to express the amplitude of the noise source (i.e.
$\overline{\eta}$) in terms of the macroscopic properties of the
system. Answering this question requieres to derive Langevin-like
equations for the hydrodynamic field of granular fluids stating from
kinetic theory or statistical mechanics descriptions.

This research was supported by the Ministerio de Educaci\'{o}n y
Ciencia (Spain) through Grant No. FIS2005-01398 (partially financed
by FEDER funds). P. Maynar's work was carried under the HPC-EUROPA
project (RII3-CT-2003-506079).


\begin{thebibliography}{}

\bibitem{LyL59} L.D. Landau and E.M. Lifshitz, {\em Fluid Mechanics}
(Pergamon Press, Oxford 1959).

\bibitem{KTyN85} R. Kubo, M. Toda, and N. Hashitsume, {\em
Statistical Physics II} (Springer-Verlag, Berlin, 1985).

\bibitem{OyS06} For a recent tutorial review see J.M. Ortiz de Z\'{a}rate
and J.V. Sengers, {\em Hydrodynamic fluctuations} (Elsevier,
Amsterdam, 2006).

\bibitem{JNyB96} H.M. Jaeger, S.R. Nagel, and R.P. Behringer, Rev.
Mod. Phys. {\bf 68}, 1259 (1996).

\bibitem{AMBLyN03} G. D'Anna, P. Mayor, A. Barrat, V. Loreto, and F.
Nori, Nature {\bf 424}, 909 (2003).

\bibitem{Ha83} P.K. Haff, J. Fluid Mech. {\bf 134}, 401 (1983).

\bibitem{vNEByO97} T.P.C. van Noije, M.H. Ernst, R. Brito, and
J.A.G. Orza, Phys. Rev. Lett. {\bf 79}, 411 (1997).

\bibitem{Go03} I. Goldhirsch, Annu. Rev. Fluid Mech. {\bf 35}, 267
(2003).

\bibitem{BDKyS98} J.J. Brey, J.W. Dufty, C.S. Kim, and A. Santos,
Phys. Rev. E, {\bf 58}, 4638 (1998).

\bibitem{GyD99} V. Garz\'{o} and J.W. Dufty, Phys. Rev. E {\bf 59},
5895 (1999).

\bibitem{Lu01} J.F. Lutsko, Phys. Rev. E {\bf 63}, 061211 (2001).

\bibitem{BRyM04} J.J. Brey, M.J. Ruiz-Montero, and F. Moreno, Phys.
Rev. E {\bf 69}, 051303 (2004).

\bibitem{BRyC99} J.J. Brey, M.J. Ruiz-Montero, and D. Cubero,
Europhys. Lett. {\bf 48}, 359 (1999).

\bibitem{BGMyR04} J.J. Brey, M.I. Garc\'{i}a de Soria, P. Maynar,
and M.J. Ruiz-Montero, Phys. Rev. E {\bf 70}, 011302 (2004).


\end{thebibliography}
\end{document}